# Elasticutor: Rapid Elasticity for Realtime Stateful Stream Processing

*Li Wang* [#], *Tom Z. J. Fu* [#], *Richard T.B. Ma* [∗], *Marianne Winslett* [§], *Zhenjie Zhang* [#]

[#] *Advanced Digital Sciences Center, Illinois at Singpore Pte. Ltd., Singapore*
{wang.li, tom.fu, zhenjie}@adsc.con.sg

[∗] *National Univeristy of Singapore, Singapore*
tbma@comp.nus.edu.sg

[§] *University of Illinois Urbana-Champaign, USA*
winslett@illinois.edu

## Abstract

Elasticity is highly desirable for stream processing systems to guarantee low latency against workload dynamics, such as surges in data arrival rate and fluctuations in data distribution. Existing systems achieve elasticity following a *resource-centric* approach that uses dynamic key partitioning across the parallel instances, i.e. *executors*, to balance the workload and scale operators. However, such operator-level key repartitioning needs global synchronization and prohibits rapid elasticity. To address this problem, we propose an *executor-centric* approach, whose core idea is to avoid operator-level key repartitioning while implementing each executor as the building block of elasticity. Following this new approach, we design the Elasticutor framework with two level of optimizations: i) a novel implementation of executors, i.e., *elastic executors*, that perform elastic multi-core execution via efficient intra-executor load balancing and executor scaling and ii) a global model-based scheduler that dynamically allocates CPU cores to executors based on the instantaneous workloads. We implemented a prototype of Elasticutor and conducted extensive experiments. Our results show that Elasticutor doubles the throughput and achieves an average processing latency up to 2 orders of magnitude lower than previous methods, for a dynamic workload of real-world applications.

## 1 Introduction

Distributed stream systems [4, 5, 1, 8] enable realtime data processing over fast-moving and continuous streams, and have been widely used in applications including fraud detection, surveillance analytics and quantitative finance. In such systems, the application logic is modeled as a graph of computation, where each vertex represents an *operator* associated with user-defined processing logic and each edge specifies the input-output relationship of data streams between the operators. To en-able large-scale data processing, the input stream to an operator is defined under a *key space* that can be partitioned into subspaces. Parallel execution instances, i.e. *executors*, are created to statically bind each key subspace to an amount of computational resource, typically a CPU core. As a result, each executor can conduct computation associated with its key subspace independently.

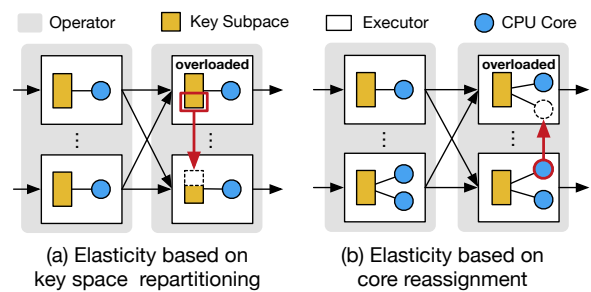

Figure 1: Comparison of elasticity mechanisms: resource-centric (left) vs. executor-centric (right).

However, severe performance degradation is observed when the application's workload fluctuates [12, 25]. From a temporal perspective, the aggregate workload fed to an operator might surge in a short period of time, making the operator a bottleneck for the entire processing pipeline, because it utilizes a fixed number of executors bound to particular computational resources. From a spatial perspective, the workload distribution over the key space might be unstable, resulting in a skewed workload across the executors of an operator with low CPU utilization in some and overload in the others. To adapt to workload fluctuation, prior work [25, 26, 12, 11] proposed solutions to enable *elasticity*, i.e., operator scaling and load balancing. All these existing solutions are *resource-centric*, in which executors are bound to particular resources and elasticity is achieved by dynamically repartitioning the keys across the executors.

Figure 1(a) illustrates a scenario where an executor is overloaded due to imbalance in workload distribution. To relieve the performance bottleneck, the key space is repartitioned such that a certain amount of workload along with the corresponding keys in the overloaded executor is migrated to a lighter-loaded executor. However, this operator-level key space repartitioning requires a time-consuming protocol [25, 12] to maintain state consistency. In particular, the system needs to sequentially perform the following operations: a) pause all the upstream executors sending tuples downstream; b) wait for all the in-flight tuples to be processed by the executors; c) migrate the state among the executors according to the new key space partitioning; and d) update the routing tables of all the upstream executors. Because both inter-operator routing update and inter-executor state migration require expensive global synchronization, the key space repartitioning typically lasts several seconds and thus introduces undesirable processing delay.

To achieve rapid elasticity, we propose a new *executor-centric* paradigm. The core idea is to statically partition the key space of an operator among its executors but dynamically assign CPU cores to each executor based on its instantaneous workload. Figure 1(b) illustrates that instead of repartitioning the key space, the new approach resolves workload imbalance by reassigning CPU cores from a lighter-loaded executor to the overloaded executor. As each executor possesses a fixed key subspace, the new approach achieves *inter-operator independence*, i.e., upstream operators do not need to synchronize with downstream ones, and *inter-executor independence*, i.e., states associated with key subspaces do not need to be migrated across executors. In other words, this new approach gracefully decouples the binding between operator-level key space repartitioning and dynamic provisioning of computational resources.

Based on the executor-centric approach, we designed the Elasticutor framework with two levels of optimization. At a global level, a model-based dynamic scheduler is designed to optimize the core-to-executor assignment based on the measured performance metrics. At the executor level, implemented as a lightweight distributed subsystem, each *elastic executor* evenly distributes its workload over its assigned CPU cores and scales efficiently when CPU cores are added to or removed from it. We have implemented a prototype of Elasticutor on Apache Storm [4] and conducted extensive experiments using both synthetic and real datasets. The results show that Elasticutor doubles the throughput and achieves orders of magnitude lower latency than existing methods.

The rest of this paper is organized as follows. Section 2 compares the executor-centric paradigm to previous approaches and gives an overview of Elasticutor. Sections 3 and 4 present the designs of elastic executors and the dynamic scheduler, respectively. Section 5 discusses experimental results. Section 6 review the related work. Section 7 concludes the paper.

## 2 Execution Paradigm and Framework

In this section, we first introduce the basic concepts of stateful stream processing, review two existing execution paradigms and propose a new executor-centric approach. We then give an overview of the Elasticutor framework.

### 2.1 Basic Concepts

We consider a real-time stateful stream processing system on a cluster of machines, called *nodes*, connected by fast network devices. A *stream* is an unbounded sequence of tuples. Tuples from the input stream(s) continuously arrive at the system and are immediately processed by the system. A user application is modeled as a directed graph of computation, called a topology, where the vertices are the operators with user-defined processing logic and the edges represent the sequence of processing among the operators. For each pair of adjacent operators, tuples of a stream are generated by the upstream operator and consumed by the downstream operator. An operator has an internal state that contains the information needed for computation and is updated during the processing of input tuples. To distribute and parallelize the computation, the state of an operator is implemented as a divisible data structure defined on a key space. The system partitions the key space into *key subspaces* and creates a parallel instance, called an executor, with identical data processing logic for each of them. To correctly route tuples to the downstream executors, routing tables are maintained in the upstream executors. Because processing the same sequence of input tuples in different orders may result in different output tuples and states, a basic requirement in stateful computation is to process the tuples of the same key in order of arrival.

Stream processing workloads are often dynamic in that the input rate to an operator and the key distribution of tuples fluctuate over time. To guarantee the performance under a dynamic workload, computational resources, i.e., CPU cores, much be appropriately provisioned to the operators so as to ensure 1) *operator scaling*, i.e., CPU cores are dynamically allocated to operators according to their workloads; and 2) *load balancing*, i.e., the workload of each operator is evenly distributed across the allocated CPU cores. Without achieving the former, some operators may be overloaded or over-provisioned, becoming a performance bottleneck or wasting computational resources, respectively. Without achieving the latter, some CPU cores will be overloaded while others will be underutilized, resulting in poor performance. We refer to the mechanism of operator scaling and load balancing as *elasticity*. To retain high performance under dynamic workloads, rapid elasticity is a crucial requirement.

| paradigms | operator-level key partitioning | CPU-to-executor assignment | elasticity |
|---|---|---|---|
| static | static | one-to-one | N/A |
| resource-centric | dynamic | one-to-one | slow |
| executor-centric | static | many-to-one | rapid |

Table 1: Comparison of three execution paradigms.

## 2.2 Three Execution Paradigms

Existing stream systems follow two paradigms: the static approach and the resource-centric approach, whose main features are summarized in Table 1.

The static approach implements each operator with a fixed number of executors and uses static operator-level key partitioning to distribute the workload among the executors. Each executor consists of a single data processing thread bound to an assigned CPU core. Due to the static key partitioning and one-to-one binding of CPU cores to executors, the static approach simplifies system implementation and is adopted in most state-of-the-art systems [4, 2]. However, since it can neither balance the workload across the allocated CPU cores nor adjust the number of CPU cores assigned to a particular operator, this approach has no elasticity and works inefficiently under a dynamic workload.

The resource-centric approach resolves the limitation of the static approach by supporting dynamic operator-level key partitioning, while following the same implementation of the executors as in the static approach. By operator-level key repartitioning, the resource-centric approach achieves elasticity, as it can migrate some keys with their corresponding workload from overloaded executors to the lighter-loaded executors to balance the workload, or from existing executors to a newly created executor to scale out an operator. However, as discussed in the introduction section, this operator-level key repartitioning is a time-consuming procedure, during which expensive global synchronization is required to migrate the state and to update the routing tables of all the upstream executors. Therefore, the resource-centric approach does not achieve rapid elasticity and can only tackle a very limited degree of workload dynamics.

To achieve rapid elasticity, we propose a new execution paradigm: *the executor-centric* approach. Our idea comes from the observation that the operator-level key repartitioning is too expensive to achieve rapid elasticity. Unlike the resource-centric approach, the executor-centric approach uses static operator-level key partitioning but implements each executor as the building block of elasticity to handle workload fluctuation. In particular, each executor is designed to utilize various amount of computation resources by creating or removing data processing threads on the fly. Therefore, to achieve load balancing and operator scaling, the system can dynam-ically assign an appropriate number of CPU cores to each elastic executor rather than performing the expensive operator-level key repartitioning. Compared with operator-level key repartitioning, reassignment of CPU cores and executor-level load balancing can be achieved efficiently, since they do not need any inter-operator or inter-executor synchronization. Fundamentally, our new approach achieves rapid elasticity by avoiding global synchronization.

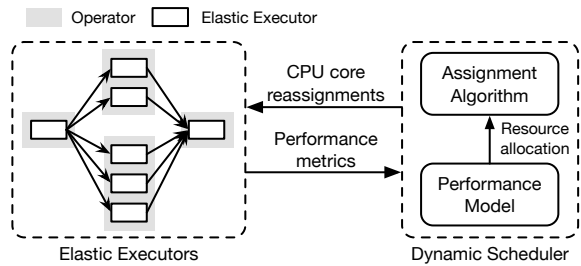

Figure 2: Overview of the Elasticutor framework.

## 2.3 Overview of Elasticutor Framework

Following the executor-centric approach, we design the Elasticutor framework that consists of elastic executors and a dynamic scheduler, as illustrated in Figure 2.

Unlike the static and resource-centric approaches, each executor now is implemented as a lightweight, self-contained, distributed subsystem, called an elastic executor, responsible for processing inputs associated with a fixed key-subspace. To adapt to the workload fluctuations, an elastic executor can utilize a dynamic number of CPU cores, possibly from multiple nodes, as determined by the dynamic scheduler. To fully utilize its allocated CPU cores in presence of key distribution fluctuation, an elastic executor has an efficient internal load balancing mechanism that evenly distributes the computation of its input stream across the allocated CPU cores. The design of elastic executors is discussed in details in Section 3.

Given the goal of guaranteeing a user-specified processing latency, the dynamic scheduler determines the desirable number of CPU cores each elastic executor should be provisioned under the instantaneous workload. It employs a performance model based on queuing networks and uses collected performance metrics of the elastic executors as inputs to generate resource allocation decisions. Based on the existing core-to-executor assignment and the availability of CPU cores in the cluster, the scheduler refines the assignment to accommodate the new resource allocation plan, while taking both the CPU reassignment overhead and the locality of computational resources into consideration. The scheduler is discussed in Section 4.

## 3 Elastic Executor

To efficiently utilize CPU resources, an elastic executor is designed to adapt to two dynamics: 1) *changes in key distribution* and 2) *CPU core reassignments*, as illustrated in Figure 3. The former results from fluctuations in the input stream, while the latter is determined by the scheduler for global performance optimization. To distribute the workload over its computational resources, an elastic executor creates a *task* for each assigned CPU core and distributes input data tuples over them. Upon a CPU reassignment, a new task will be created or an existing task will be deleted. Both changes in key distribution and CPU core reassignments introduce unbalanced workload among the tasks, resulting in resource underutilization or performance degradation. Therefore, a central design question is *how to evenly distribute the workload among the tasks in presence of such dynamics.*

In the rest of this section, we first discuss the intra-executor load balancing policy and then describe an implementation that enables highly efficient workload redistribution with state consistency.

### 3.1 Intra-Executor Load Balancing

Within each elastic executor, executor-level key space repartitioning is dynamically performed to balance the workload across its tasks. In what follows, we first discuss the granularity of key space partitioning that makes trade-offs between the maintenance overhead and the quality of load balancing, and then present the load balancing algorithm that minimizes the state migration overhead associated with key reassignments.

A straightforward way of achieving load balancing is to monitor the workload for each key and reassign keys from overloaded tasks to underutilized ones. However, for applications with very large key spaces, this fine-grained method suffers from high memory consumption, since it needs to maintain the assigned task ID and the workload statistics for every single key. To reduce the maintenance overhead, we balance the workload in a coarser-grain rather than on a per-key basis. Specifically, we statically partition keys into mini-partitions, called *shards,* using a hash function and dynamically assign shards to tasks. The choice of the number of shards provides trade-offs between the quality of load balancing and maintenance overhead. With more shards, the most frequent keys are more likely to be hashed to different shards and can be assigned to different tasks for better load balancing; however, too many shards will lead to over-sized routing tables and high overhead for maintaining the statistics. The appropriate choices for the number of shards will be discussed in Section 5.3.

To guarantee state consistency during load balancing,

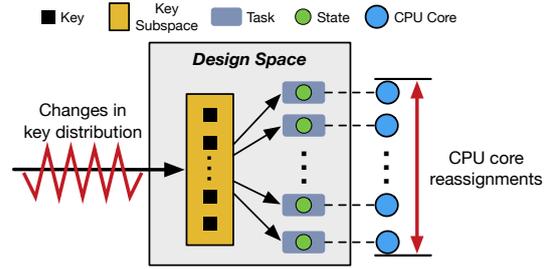

Figure 3: The design space of an elastic executor against changes in key distribution and core reassignment.

states have to be migrated along with their associated shards among the tasks, leading to migration overhead and delay. Consequently, for rapid load balancing, the number of reassigned shards should be minimized. This optimization problem can be interpreted as a multi-way partitioning problem [18], which is known to be NP-hard. We use a heuristic algorithm similar to the First-Fit-Decreasing algorithm [14] to solve it. Our intra-executor load balancing algorithm refines the shard-to-task assignment in rounds until the workload imbalance factor $\delta$ is below a predefined threshold $\theta$. In each round, among all the possible reassignments that reassign a shard from the most overloaded task to the least loaded task, the algorithm picks the shard reassignment that reduces $\delta$ the most. In out implementation, we define $\delta$ as the the ratio of maximum task workload to the average workload of all the tasks, and choose $\theta = 1.2$, allowing a maximum imbalance of 20% deviation from the average workload of the tasks.

### 3.2 Executor Components

An elastic executor can utilize computational resources on multiple physical nodes, and is implemented as a lightweight and self-contained distributed subsystem, as illustrated in Figure 4. Each elastic executor primarily resides in one physical node, called its *local node*, where it runs a local *main process* to host auxiliary daemon threads such as the receiver and emitter and in-memory structures like the routing table. For each allocated CPU core, a task, implemented as a data processing thread, is created in the process. For performance enhancement, each task maintains a pending queue to buffer its unprocessed input tuples. To utilize CPU cores on a remote node, a remote process can be created to host remote tasks for remote data processing.

Following the executor-centric approach, each elastic executor owns a private key subspace and maintains the states associated with its key subspace. We employ a two-tier design, implemented in the routing table as shown in the central rectangle in Figure 4, to map each input tuple to its designated task based on the load-balancing algorithm described in Section 3.1. In particular, the first tier statically partitions the key subspace into shards using a hash function; the second tier explicitly maintains a dynamic shard-to-task mapping, which gets updated upon shard reassignments.

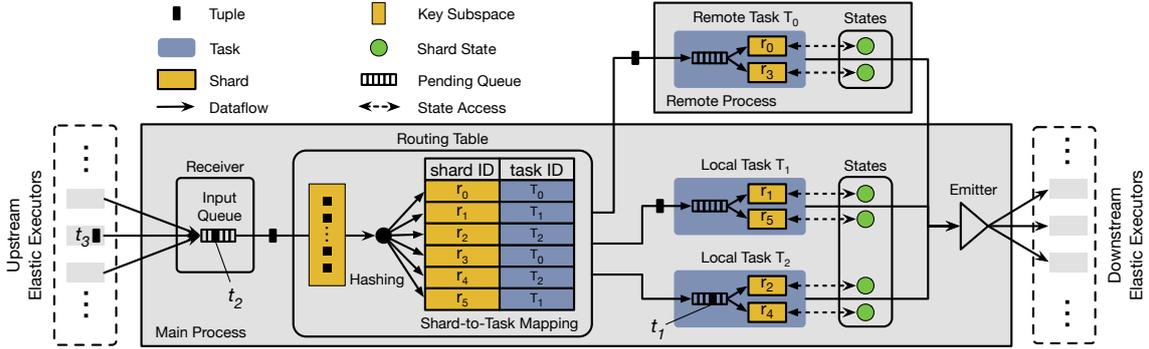

Figure 4: The internal structures and working mechanisms of an elastic executor.

During the reassignment of a shard, the state of the shard needs to be migrated to a new task, possibly in a remote node. For ease of management, external distributed key-value store, such as RAMCloud [22], can be used to provide a unified state access interface to all tasks, thus avoiding the necessity of state migration in shard reassignments. However, this method sacrifices the efficiency of task execution, because accessing states in external storage requires state serialization and network transfer, which introduces undesirable delay. To enable efficient state access, existing systems [4, 2] often allow each task to maintain its states as a private data structure, through which direct and efficient state access is enabled. However, whenever a shard is assigned to a new task, inter-task communications are needed to migrate the state of the shard between tasks.

To minimize the state migration overhead and guarantee the state access efficiency simultaneously, we employ an *intra-process state sharing* mechanism in the elastic executors. To be more specific, each process of an elastic executor maintains the states of its tasks in a lightweight in-memory key-value store and provides a *state access interface* to its tasks for state reads and updates on a per-key basis. While retaining efficient state access performance, this design avoids state migration when shards are reassigned between tasks on the same node, because the newly assigned task can always access the shard's state via the interface without state migration. Given the increasing number of CPU cores on modern processors, many tasks can be created on a single node. Consequently, intra-process state sharing can significantly reduce shard reassignment overhead. Furthermore, our dynamic scheduler also optimizes the locality of CPU resources for the elastic executors, providing the executors more opportunities to benefit from state sharing.

### 3.3 Consistent Shard Reassignment

Although state sharing improves the efficiency of shard reassignment, special attention needs be paid to guarantee state consistency. Generally speaking, despite using the similar procedure as in key repartitioning of the resource-centric approach, we achieve efficient shard reassignment with state consistency by taking advantage of the inter-operator and inter-executor independence enabled by the executor-centric approach.

Consider the case illustrated in Figure 4, where a tuple $t_1$ is in the pending queue of task $T_2$, a tuple $t_2$ just arrived at the entrance of the executor's main process, and a tuple $t_3$ is to be emitted by an upstream executor. Suppose all three tuples belong to shard $r_4$. If shard $r_4$ is reassigned from the *source task $T_2$* to a new destination task before $t_1$ is processed or before the routing of $t_2$ and $t_3$ are updated, the state will become inconsistent. In particular, if the destination task is local, e.g., $T_1$, then $t_2$ might be processed before $t_1$, violating the order preserving requirement. If the destination task is remote, e.g., $T_0$, the modifications to states made by $t_1$ will be lost.

**Inter-Operator Consistent Routing:** To guarantee consistent routing of tuples, e.g., $t_3$, from upstream operators to the correct processes where the assigned tasks reside, an elastic executor implements a *receiver* daemon in its local main process as the single entrance for all tuples coming from upstream operators. The receiver routes tuples to the appropriate tasks, local or remote, based on the internal routing table. Similarly, an *emitter* daemon is implemented in the main process as the single exit of the executor to forward output tuples generated by the tasks to downstream operators. Remote processes only communicate with the receiver and the emitter on the main process of the elastic executor. Therefore, regardless of how shards are reassigned among the tasks within an elastic executor, upstream and downstream operators always send tuples to or receive tuples from the executor via its receiver and emitter, avoiding any inter-operator synchronization caused by shard reassignments. In con-trast, the resource-centric approach redistributes workload by operator-level key space repartitioning, leading to synchronization with all the upstream executors.

Note that compared with the resource-centric approach where tuples from upstream executors are directly routed to the processing threads in the downstream operator, Elasticutor may involve additional remote data transfer between the receiver/emitter and the remote tasks. This is the trade-off we make to achieve rapid elasticity. In typical workloads, the remote data transfer is not the performance bottleneck, as shown in Figure 5.2. In Section 5.3, we discuss how to avoid/reduce remote data transfer in some extreme workloads by properly configuring the number of executors of an operator.

**Intra-Executor State Consistency:** To guarantee state consistency during the reassignment of a shard, an elastic executor employs a procedure similar to the operator-level key space repartitioning used in the resource-centric approach, but does not involve any global synchronization. The key is to ensure that the pending tuples, i.e., the unprocessed tuples of the shard queued in the source task, have been processed before the shard state is migrated to the destination task. During the reassignment of shard $r_4$ in Figure 4, the routing for tuples of $r_4$ is paused and a labeling tuple is sent to its source task $T_2$. Since tasks process their input tuples on a first-come-first-served basis, any pending tuple already sent to $T_2$ is guaranteed to be processed when $T_2$ pulls the labeling tuple from its pending queue. After that, the state of $r_4$ is migrated to the destination task. State migration is omitted if the shard is reassigned to a task local to its source task. After the state migration, the shard-to-task mapping is updated in the routing table before the routing for tuples of $r_4$ is resumed.

## 4 Dynamic Scheduler

The objective of the dynamic scheduler is to satisfy user-defined latency requirements by adaptively allocating CPU cores to the elastic executors under a changing workload. By using instantaneous performance metrics measured by the system, the scheduler first estimates the number of cores needed for each executor based on a queueing network model, and further (re)assigns the physical cores to the executors so as to minimize the reallocation overhead and maximize the locality of computation within the executors.

### 4.1 Model-Based Resource Allocation

We model a topology $\mathscr{E} = \{1, \cdots, m\}$ of $m$ elastic executors as a Jackson network, in which each executor $j \in \mathscr{E}$ is regarded as an M/M/$k_j$ system [27], where $k_j$ denotes the number of allocated CPU cores to $j$. The average processing latency of an input stream, denoted as $\mathbb{E}[T]$, can be calculated as a function of the resource allocation decision $\mathbf{k}$ as

$$\mathbb{E}[T](\mathbf{k}) = \frac{1}{\lambda_0} \sum_{j=1}^{m} \lambda_j \mathbb{E}[T_j](k_j), \quad (1)$$

where $\lambda_0$ denotes the arrival rate of the input stream, and $T_j$ and $\lambda_j$ denote the average processing time and the arrival rate of executor $j$, respectively. Each $\mathbb{E}[T_j](k_j)$ is bounded when $k_j > \lambda_j/\mu_j$, where $\mu_j$ denotes the processing rate of elastic executor $j$ and can be calculated as a function of the parameters $\lambda_0$, $\{\lambda_j\}$ and $\{\mu_j\}$ measured by the system. Based on Equation (1), the scheduler attempts to find an allocation $\mathbf{k}$ to ensure that $\mathbb{E}[T]$ is no larger than the user-specified latency target $T_{max}$, while minimizing the total number of CPU cores, i.e., $\sum k_j$. In particular, each $k_j$ is initialized to be $\lfloor \lambda_j/\mu_j \rfloor + 1$, which is the minimal requirement to make the system stable. We repeatedly add 1 to the value in the vector $\mathbf{k}$ that leads to the most significant decrease in $\mathbb{E}[T]$, until $\mathbb{E}[T] \leq T_{max}$ or $\sum k_j$ exceeds the number of available CPU resources. This greedy algorithm has shown to be optimal [15] in finding the solution $\mathbf{k}$.

### 4.2 CPU-to-Executor Assignment

The performance model only suggests the number of CPU cores needed; the scheduler still needs to determine the mapping from the physical CPU cores to the executors. To accommodate a new allocation resulting from workload fluctuation, the scheduler may need to update the existing core-to-executor assignment. In a typical reassignment, workload needs to be redistributed to the newly assigned cores, possibly remote from the executor's local node. To achieve this, the states associated with the workload have to be migrated and future data transfers between the receiver/emitter and the newly assigned cores are introduced. Because the CPU assignment determines the locations of the reassigned cores and the executors involved, it influences 1) the state migration costs during the transition, and 2) the remote data transfer costs afterwards. To optimize execution efficiency, we search for CPU-to-executor assignments that minimize migration costs, while constraining the computation locality to limit future remote data transfer costs.

To model the migration costs, we consider a cluster of $n$ nodes where each node $i$ has $c_i$ CPU cores. For any executor $j \in \mathscr{E}$, we denote the node where its main process resides by $I(j)$ and the number of cores assigned to it on all nodes by a column vector $\mathbf{x_j} = (x_{1j}, \cdots, x_{nj})^T$. We define $X_j = \sum_{i=1}^{n} x_{ij}$ as the total number of assigned cores for $j$ and denote a CPU-to-executor assignment by a matrix $X = (\mathbf{x_1}, \cdots, \mathbf{x_m})$. Given any new allocation $\mathbf{k}$, a transition from an existing assignment $\tilde{X}$ to a new assignment $X$ needs to perform a set of CPU allocations/deallocations. The overhead of core reassignment is dominated by the state migration cost, which is proportional to the size of state moved across the network. We denote the aggregate state size of any executor $j$ by $s_j$. For simplicity, we assume the shards of an elastic executor are evenly distributed across the allocated CPU cores; and therefore, the amount of state data associated with each CPU core is approximately $s_j/X_j$. Consequently, we can estimate the cost of transition from an existing assignment $\tilde{X}$ to a new assignment $X$ as $C(X|\tilde{X}) = \sum_{j=1}^{m}\sum_{i=1}^{n} \max(0, \frac{s_j \tilde{x}_{ij}}{\tilde{X}_j} - \frac{s_j x_{ij}}{X_j})$, where each term in the summation measures the cost for executor $j$ to migrate its state out of node $i$. Given any allocation $\mathbf{k}$, available cores $\mathbf{c}$ and an existing assignment $\tilde{X}$, we formulate the CPU assignment problem as follows.

$$\begin{aligned}
\underset{X}{\text{minimize}} \quad & C(X|\tilde{X}) \\
\text{s.t.} \quad & \text{(a)} \sum_{j=1}^{m} x_{ij} \leq c_i, && \forall i \leq n; \\
& \text{(b)} \ X_j \geq k_j, && \forall j \in \mathscr{E}; \\
& \text{(c)} \ x_{I(j)j} = X_j, && \forall j \in \mathscr{E}(\varphi).
\end{aligned} \quad (2)$$

The above optimization problem minimizes the migration costs $C(X|\tilde{X})$, subject to (a) the capacity of CPU cores, (b) allocation requirement constraint and (c) *computation locality* constraint, i.e., requiring all cores assigned to the set $\mathscr{E}(\varphi)$ of executors to be on their local nodes. The system measures the instantaneous per-core data-intensity of any executor $j$ by its total input and output data rates divided by the number of cores $k_j$, and $\mathscr{E}(\varphi)$ denotes the set of executors whose data-intensity is above a threshold $\varphi$. Because data-intensive executors will incur higher network costs if their assigned cores are remote, we enforce the computation locality by avoiding assigning remote cores to members of $\mathscr{E}(\varphi)$. This integer programming problem can be reduced to the NP-hard multiprocessor scheduling problem [16]. Thus, we design an efficient greedy Algorithm 1 to find an approximate solution. For any assignment $X$, we define $\mathscr{E}^+ = \{j \in \mathscr{E} | X_j < k_j\}$ to be the set of under-provisioned executors, $\mathscr{E}_\Delta^+ = \{j \in \mathscr{E}^+ \cap \mathscr{E}(\varphi)\}$ to be the subset of data-intensive executors, and $\mathscr{E}^- = \{j \in \mathscr{E} | X_j > k_j\}$ to be the set of over-provisioned executors. We use $C_{ij}^+(X)$ and $C_{ij}^-(X)$ to denote the overhead of allocating/deallocating a CPU core on node $i$ to/from executor $j$, respectively, which can be derived as $C_{ij}^+(X) = s_j(X_j - x_{ij})/(X_j(X_j+1))$ and $C_{ij}^-(X) = s_j(X_j - x_{ij})/(X_j(X_j-1))$.

Algorithm 1 sorts the executors in $\mathscr{E}^+$ by data-intensity in descending order and tries to assign the target number of CPU cores to each executor $j$ one by one by deallocating cores from other executors. Specifically, if elastic executor $j$ is data-intensive, i.e, $j \in \mathscr{E}(\varphi)$, it

---

**Algorithm 1:** Dynamic Allocation Algorithm

**Input**: allocation $\mathbf{k}$, assignment $\tilde{X}$, CPU cores $\mathbf{c}$, threshold $\varphi$
**Output**: new assignment $X$

1 Initialize the new partitioning as $X = \tilde{X}$;
2 Find the under- and over-provisioned executors $\mathscr{E}^+$ and $\mathscr{E}^-$, and the data-intensive executors $\mathscr{E}_\Delta^+$;
3 Sort $\mathscr{E}^+$ based on the data-intensity of the executors;
4 **for** *each $j \in \mathscr{E}^+$ in non-descending order* **do**
5     **while** *CPU cores are insufficient, i.e., $X_j < k_j$* **do**
6        **if** *$j$ is data-intensive, i.e., $j \in \mathscr{E}(\varphi)$* **then**
7           $i = I(j); \quad j = \underset{j \in \mathscr{E}^- \mathscr{E}_\Delta^+}{\arg\min} C_{ij}^-(X)$
       **else**
          $(i,j) = \underset{j \in \mathscr{E}^-, 1 \leq i \leq n}{\arg\min} C_{ij}^-(X) + C_{ij}^+(X)$
1        **if** *$(i,j)$ is found* **then**
11           $x_{ij} = x_{ij} - 1; \quad x_{ij} = x_{ij} + 1$
12        **else**
13           return FAIL;

14 return $X$;

---

only accepts CPU cores on node $i = I(j)$, to avoid creating remote tasks. Consequently, among all the non-data-intensive executors, the algorithm finds a CPU core on node $I(j)$ that can be reassigned to $j$ with minimal deallocation overhead (Line 7). In contrast, if $j$ is not data-intensive, it accepts CPU cores on any node. The algorithm searches all the executors in $\mathscr{E}^-$ for an executor with a CPU core that can be reassigned to $j$ with the minimal deallocation and allocation overhead (Line 9). In either case, if such a valid core reassignment is found, the algorithm added it to the new assignment $X$; otherwise, it returns FAIL, which indicates that no feasible solution can be found and implies that a higher data-insensitivity threshold $\varphi$ is required to obtain a feasible solution.

The choices of $\varphi$ provide trade-offs between the feasibility of Equation 2 and the computation locality of the elastic executors. Since the dynamic assignment algorithm is very efficient, we run the algorithm using a low default value $\varphi = \tilde{\varphi}$. If no feasible solution is found, we double $\varphi$ and re-run the algorithm until we find one. In our experiments, we set $\tilde{\varphi}$ to be 512 KB/s, below which the benefit of computation locality is negligible.

Although our scheduling algorithm improves computation locality effectively, it is possible that in some extreme workloads, e.g., highly skewed key distribution, some executors may run excessive tasks, thus introducing extensive remote data transfer. To tackle this problem, we can detect and split those overloaded executors at a coarse time granularity, e.g., every 10 minutes. This is also useful when system workload has increased so much that the system needs to gracefully scale out to much more nodes, e.g., from initial 10 nodes to 100 nodes. Similarity, when the total workload decreases substantially, it is desirable to merge some idle executors so that some nodes can be freed up. In the future work, we plan to design a hybrid framework that uses elastic executors to provide rapid elasticity and infrequently performs operator-level key space repartitioning for long-term optimizations, such as resolving an overloaded executor or scaling out/in the entire system.

## 5 Performance Evaluation

We implemented a prototype of Elasticutor in about 10,000 lines of Java on Apache Storm [4], a state-of-the-art open-source stream processing system. Storm follows the static approach and its operators are implemented by users via an abstract class, *Bolt*. We added a new abstract class, *ElasticBolt*, which provides the same programming interface as Bolt, but exposes a new state access interface to the user space. For any operator defined as an ElasticBolt, Elasticutor creates a number of elastic executors with built-in state management, metrics measurement and elasticity functionalities. The dynamic scheduler is implemented as a daemon process running on Storm's master node (nimbus). We compare the performance of Elasticutor with that of the static approach (the default Storm) and resource-centric (RC) approaches. We implemented RC based on Storm by enabling creation/deletion of executors and operator-level key repartitioning. For fair comparison, RC uses the same performance model, load balancing algorithm and intra-process state sharing mechanism as Elasticutor.

Our experiments are conducted on EC2 with 32 t2.2xlarge instances (nodes), each with 8 CPU cores and 32 GB RAM running Ubuntu 16.04. The network is 1Gbps Ethernet. The executors are assigned to the nodes in a round-robin manner under all approaches. Unless otherwise stated, Elasticutor uses 32 elastic executors per operator and 256 shards per executor (8192 shards per operator). For fair comparison, we create enough executors for the operators in the static approach to fully utilize all CPU cores in the cluster; and the granularity of the key space repartitioning in the RC approach is 8192 shards per operator, the same as in Elasticutor.

### 5.1 Micro-Benchmarking

In this subsection, we use a simple yet representative topology, shown in Figure 5, which allows easy control over the workload characteristics, such as input rates and data distribution. Unless otherwise stated, each tuple consists of an integer key and a 128-byte payload, and takes an average CPU cost of 1 ms for processing. The key space contains 10K distinct values, whose frequencies follow a zipf distribution [23] with a skew factor of 0.5. The shard state is 32KB in size. To emulate workload dynamics, we shuffle the frequencies of tuple keys by applying a random permutation $\omega$ times per minutes.

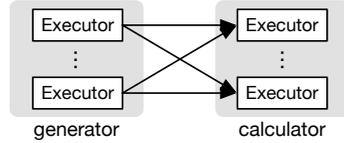

Figure 5: The micro-benchmarking simulation topology.

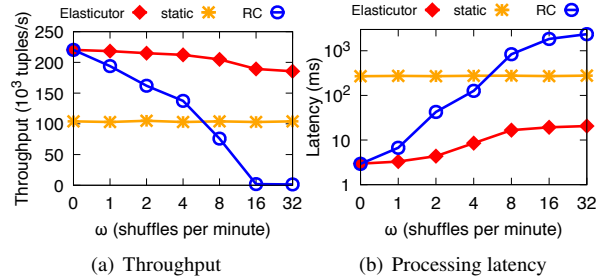

(a) Throughput  (b) Processing latency

Figure 6: Performance comparison with varying workload dynamics.

**Robustness to workload dynamics:** Figure 6 plots the throughput and average processing latency under the three approaches as $\omega$ varies along the x-axis. We observe that Elasticutor consistently outperforms the others in terms of both metrics when the workload is dynamic, i.e., $\omega > 0$. Specifically, the performance of the static approach is poor due to workload imbalanced caused by skewed key distribution, but is relatively stable across all scenarios as no elasticity operations are performed. Since both RC and Elasticutor are able to adapt to skewed key distribution, they outperform the static considerably when $\omega$ is small. However, as $\omega$ increases, although the performance of both RC and Elasticutor decreases due to higher operational costs for elasticity, the performance degradation of Elasticutor is marginal, while that of RC becomes 2-3 orders of magnitude larger, making RC useless as $\omega$ reaches 16.

To better explain the performance of the three approaches as $\omega$ varies, we focus on the scenario of $\omega = 2$, i.e., shuffle every 30 seconds, and plot the instantaneous throughput, measured in a sliding time window of 1 second, in Figure 7. We observe that the throughput of the static approach is consistently much lower than that of RC and Elasticutor, although it does not vary much. Both RC and Elasticutor exhibit a transient throughput degradation every 30 seconds, due to the executions of elasticity operations triggered by key shuffles. However, the degradation in RC is much worse and its transient period lasts 10 to 20 seconds, while that of Elasticutor only lasts 1 to 3 seconds. This explains the reason behind the widening performance gap in the two approaches as the workload becomes more dynamic.

**Shard reassignment cost:** Because both the RC approach and Elasticutor use shard reassignment to balance the workload, we compare their costs to better understand the different delays incurred. Figure 8 shows the average intra- and inter-node reassignment time per shard, broken down into synchronization time and state migration time. We observe that the shard reassignment time is much longer in RC than in Elasticutor, mainly due to the extremely long synchronization time in the RC approach. We can also see that Elasticutor takes shorter time in state migration than RC, but the difference between the two approaches in state migration is minor compared to that in synchronization time.

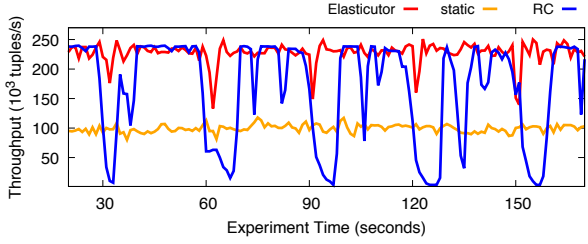

Figure 7: Instantaneous throughput with $\omega = 2$.

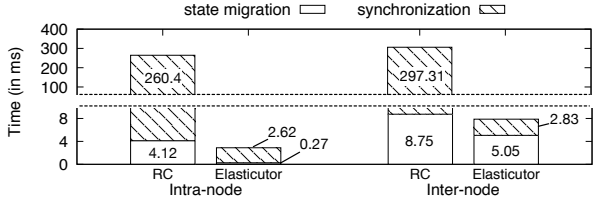

Figure 8: Breakdown of shard reassignment time.

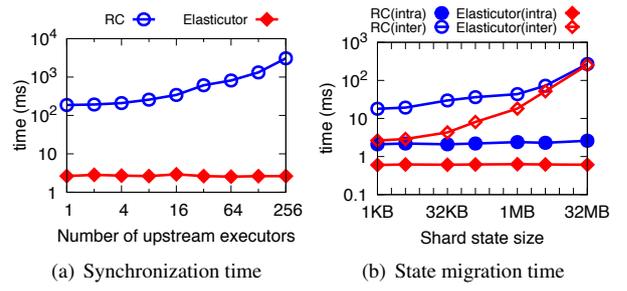

(a) Synchronization time  (b) State migration time

Figure 9: Effect of the number of upstream executors and the state size.

To gain insights into the synchronization time differences between the two approaches, we vary the number of upstream executors and show the times in Figure 9(a). We observe that the RC approach takes 2-3 orders of magnitude longer to synchronize than Elasticutor and the difference widens as the number of upstream executors increases. Elasticutor follows the resource-centric paradigm, the inter-operator independence of which makes shard reassignment a local operation within the executor, avoiding any synchronization with upstream executors. As a result, the synchronization time is around 2 ms regardless of the number of upstream executors. In contrast, RC needs to synchronize with all the upstream executors, and consequently the synchronization time is much higher and grows significantly with the number of upstream executors.

Figure 9(b) plots the state migration times as the state size varies. We observe that the latency of intra-node state migration is negligible in both approaches, because of the intra-process state sharing mechanism. The time of inter-node state migration increases significantly as the state size reaches 32 MB, where network data transfer of the state is the dominant overhead in the state migration process. The figure also shows that given the same state size, the Elasticutor takes slightly shorter time to migrate the state than RC, due to inter-executor independence enabled by the executor-centric paradigm.

## 5.2 Scalability of Elasticutor

The major advantage of Elasticutor is that it handles workload dynamics by allocating more CPU cores rather than operator-level key space repartitioning. Although in a reasonable setting an operator typically has enough executors to amortize its workload, it is still possible that a single executor may be so heavy that many remote tasks are needed, due to skewed key distribution, improper operator-level partitioning or unnecessarily few executors. Consequently, for robustness of Elasticutor, it is crucial that an elastic executor has good *scalability*, i.e., being able to efficiently scale out to many CPU cores, and does not introduce noticeable processing latency in running remote tasks. To evaluate to what extend the elastic executor can efficiently scale out, we set up only ONE elastic executor for the calculator operator, but gradually allocate more CPU cores and measure its throughput and processing latency. As each node has 8 CPU cores, the first 8 cores allocated are local, with the subsequent ones being remote. In our evaluation, we vary data intensity and operational cost of elasticity, which are the major factors affecting the scalability. The former decides the long-term cost of remote data transfer in running a remote task, and is proportional to tuple size and reversely proportional to the computational cost per tuple. The latter affects the short-term transit overhead in performing elasticity operations, and has positive correlation with the state size and workload dynamics ($\omega$).

Figure 10 plots the scalability of an executor under different computational costs (left) and tuple sizes (right). We observe that the single elastic executor generally can efficiently scale out to the whole cluster (256 CPU cores), indicating that cost of remote data transfer is negligible. We also observer that the elastic executor cannot efficiently utilize more than 16 CPU cores with a very large tuple size, e.g., 8KB, or very low computation cost, e.g., 0.01ms per tuple, indicating that the huge remote data transfer linked to the high data intensity prevents the executor from scaling. Figure 11 shows the 99% percentile latency as an elastic executor scales out. We can see that in most settings, processing latency does not increase noticeably as the elastic executor scales out, due to the efficient network data transfer enabled by Netty [3]. However, in the data-intensive workload, e.g., computational cost $\leq$ 0.1ms or tuple size $\geq$ 2KB, the latency increases greatly as the number of allocated CPU cores exceeds the points where remote data transfer becomes the performance bottleneck. Note that the latency does not grow infinitely, due to the back-pressure mechanism.

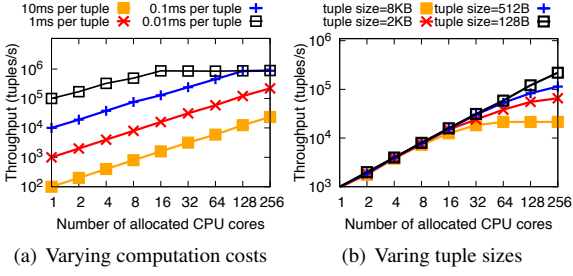

(a) Varying computation costs
(b) Varing tuple sizes

Figure 10: The scalability of a single elastic executor as data intensity varies.

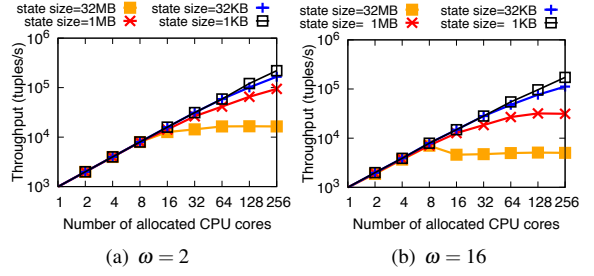

(a) $\omega = 2$
(b) $\omega = 16$

Figure 12: Throughput of a single elastic executor as operational cost of elasticity varies.

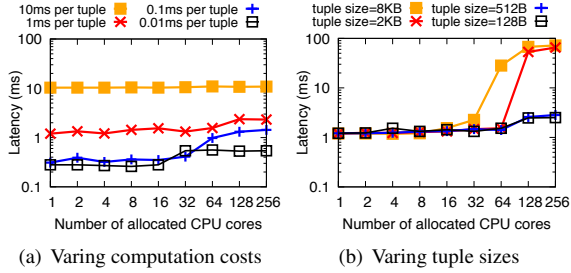

(a) Varing computation costs
(b) Varing tuple sizes

Figure 11: The 99% percentile latency as an elastic executor scales out.

Figure 12 shows the scalability of an elastic executor under various shard state sizes with $\omega = 2$ (left) and 16 (right). The results show that the elastic executor scales efficiently under all the shard state sizes but 32MB. With a large state size, the state migration becomes a performance bottleneck, which prevents the executor from efficiently using remote CPU cores. By comparing both sub-figures, we observe that as the workload dynamic $\omega$ increases to 16, the scalability under the large state size decreases considerably, due to the increased requirement of state migration linked to higher workload dynamics.

### 5.3 Choosing Appropriate Parameters

We need to determine two system parameters: the number of shards per executor, denoted as $z$, and the number of executors per operator, denoted as $y$. We used the default values of $(y,z) = (32,256)$ in our evaluations. In what follows, we evaluate their impact on system performance so as to understand how to choose appropriate parameters in practice. To make comprehensive observation, we use three representative workloads, namely the *default* workload, *data-intensive* workload and *highly dynamic* workload. Let $s$ and $\omega$ denote the tuple size in bytes and key shuffles per minute, respectively. In the default workload, $(s,\omega) = (128,2)$. We get data-intensive

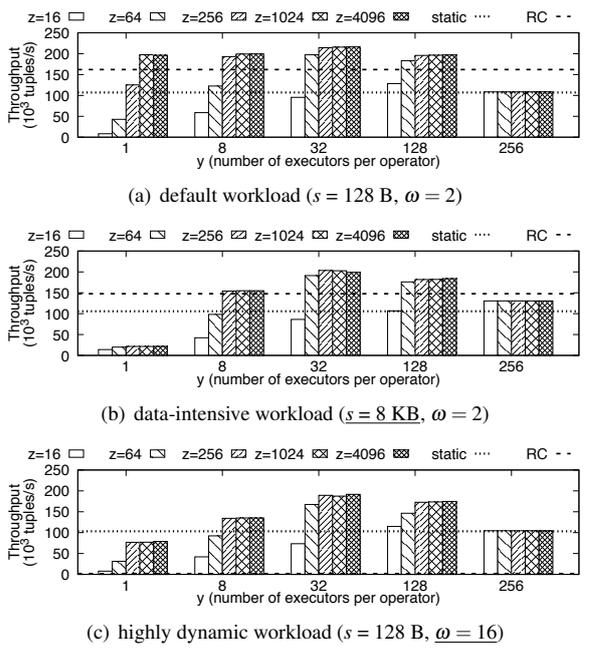

(a) default workload ($s$ = 128 B, $\omega = 2$)

(b) data-intensive workload ($s$ = 8 KB, $\omega = 2$)

(c) highly dynamic workload ($s$ = 128 B, $\underline{\omega = 16}$)

Figure 13: The impact of number of executors ($y$) and number of shards ($z$) on the throughput of Elasticutor.

workload and highly dynamic workload by increasing $s$ to 8192 and $\omega$ to 16, respectively. Figure 13 shows the system throughput with various $y$ and $z$ under the three workloads. For comparison, we also show the throughput of the static and RC approaches in the figures.

**Number of shards:** From Figure 13, we observer that as $z$ increases, the throughput generally increases though the marginal increase is diminishing. This shows when using too few shards, poor quality of intra-executor load balancing prevents elastic executors from efficiently utilizing multiple cores; however, too fine-grained sharding does not further improve throughput as intra-executor load balancing is already effective.

**Number of executors:** As shown in Figure 13(a), for a sufficiently large $z$, Elasticutor achieves promising performance except for $y$ = 256. When $y$ = 256, i.e., the number of CPU cores in the cluster, each elastic executor can only be allocated one CPU core. As such, executors lose elasticity and Elasticutor is downgraded to the static approach. By comparing Figure 13(a) with Figure 13(b), we can see that as tuple size increases to 8192, the performance of the static and the RC does not change much, while that of Elasticutor under $y = 1$ drops severely. Compared with the default workload, the cost of remote data transfer in running a remote task in the data-intensity workload is 64 times higher. This limits the scalability of a single executor and thus results in poor performance for small $y$ where a single executor needs to scale to many remote CPU cores. By comparing Figure 13(a) to Figure 13(c), we observe that as the shuffle frequency increased from 2 to 16, although the throughput decreases in general, the reduction is much greater when $y$ is small, i.e., 1 or 8. Under a dynamic workload with frequent shuffles, e.g., $\omega = 16$, more shards need to be reassigned for load balancing, incurring high migration cost. In contrast, when $y$ is sufficiently large, most executors can scale using local CPU cores and thus avoid state migration due to intra-processing state sharing mechanism; and therefore, the throughput does not decrease much. In conclusion, setting one or two executors per node is robust to various workloads.

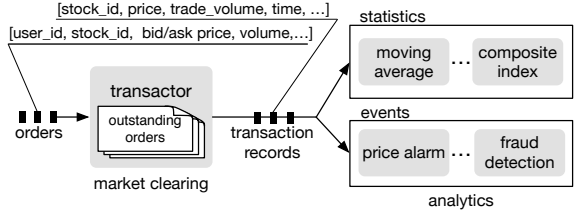

Figure 14: The topology of the SSE application.

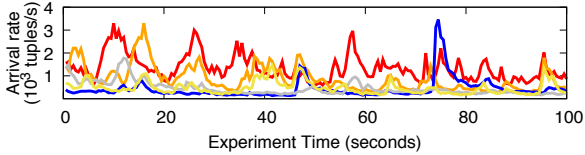

Figure 15: Arrival rates of 5 most popular stocks.

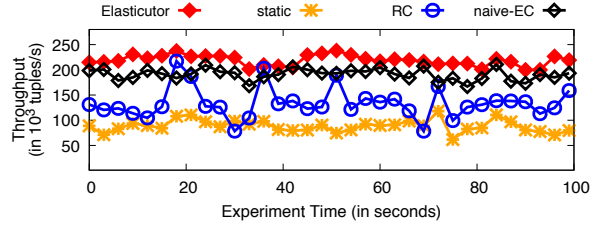

(a) Throughput

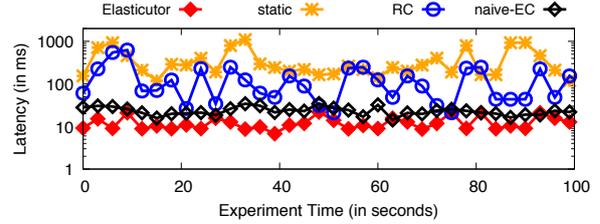

(b) Processing latency

Figure 16: Performance comparison.

## 5.4 Evaluation of Realtime Application

To evaluate the performance of Elasticutor for practical applications, we use a dataset of anonymized orders for stocks traded in the Shanghai Stock Exchange (SSE), collected over three months with around 8 million records per trading hour. The application performs the market clearing mechanism of the stock exchange and provides real-time analytics. The topology of the application is shown in Figure 14. The input stream consists of limit orders from buyers and sellers, who specify their bid and ask prices for a particular volume of a particular stock. An order tuple is 96 bytes in size. Upon the arrival of a new order, a *transactor* operator executes it against the outstanding orders and determines the quantities traded and the cash transfers made. Once such a transaction is made, a 160-byte transaction record, including the time, number of shares and price of the transaction and IDs of the seller, buyer and stock, is sent to the downstream operators, including 6 operators for statistics and 5 operators for event processing. The analytics operators generate statistics, such as the moving averages and the composite index, and trigger user-defined events, such as alarms when the transaction price of a particular stock exceeds a predefined threshold. As transactions and analytics concern individual stocks, we partition the space of stock IDs for parallel processing. Due to the unpredictable nature of stock trading, both the arrival rates and distribution of the orders of stocks fluctuate greatly over time, resulting in a highly dynamic workload. To illustrate the workload dynamics, Figure 15 shows the arrival rate of 5 most popular stocks with time.

Besides the static, RC and Elasticutor, we test a naive executor-centric (naive-EC) implementation, which is the same as Elasticutor except that optimizations for migration cost and computation locality are disabled in the scheduler. Figure 16 plots the instantaneous throughput and average processing latency under the four approaches running on 32 nodes. We observe that both naive-EC and Elasticutor outperform the static and RC approaches, approximately doubling the throughput and reducing the latency by 1-2 orders of magnitude. Although the performance gaps between the naive-EC and Elasticutor are recognizable, they are small compared to those between the executor-centric approaches and the other two approaches. This observation indicates that despite the considerable performance improvement enabled by the optimizations in the dynamic scheduler, the better performance of Elasticutor is mainly due to the advantageous executor-centric paradigm employed.

To further pinpoint the reason behind the performance gap between the naive-EC and Elasticutor, we show their state migration rate and the remote data transfer rate in Table 2. The former rate is the aggregated size of state the whole system migrates across network in a unit of time. The latter rate is the aggregated amount of data transferred in a unit of time between all the elastic executors and their remote tasks. We observe that the rates of state migration and remote data transfer under naive-EC are 5x and 10x higher than those under Elasticutor, respectively. With less state migration, it will be more efficient for the elastic executors to transition to a new resource allocation plan, thus achieving higher performance. Similarly, with less remote data transfer, more network bandwidth can be used by inter-operator data transfer, further improving the performance.

| Metrics | naive-EC | Elasticutor |
|---|---|---|
| State migration rate (MB/s) | 13.9 | 2.4 |
| Remote data transfer rate (MB/s) | 235.3 | 21.6 |

Table 2: Comparison between naive-EC and Elasticutor.

| number of nodes in the cluster | 8 | 16 | 32 |
|---|---|---|---|
| throughput ($10^3$ tuples/s) | 66.6 | 121.3 | 218.6 |
| scheduling time (ms) | 4.1 | 5.2 | 5.7 |

Table 3: Throughput and scheduling time of Elasticutor.

Finally, we evaluate the scalability of Elasticutor under the SSE workload. We vary the size of the computing cluster, i.e., the number of nodes, and measure Elasticutor's throughput and scheduling cost, i.e., the average time needed for the dynamic scheduler to calculate a new CPU-to-executor assignment. Keeping a low scheduling cost is important for the system to be adaptive to a dynamic workload. Table 3 shows the throughput and the scheduling cost as the scale increases. We observe that the throughput grows nearly linearly as the cluster grows; and the scheduling cost is around several milliseconds and grows slightly with the number of nodes.

## 6 Related Work

**Stream Processing Systems.** Early stream processing systems, such as Aurora [6], Borealis [7], TelegraphCQ [13] and STREAM [10] were designed to process massive data updates by exploiting distributed but static computational resources. With cloud computing technologies, a new generation of stream systems emerged, with emphasis on parallel data processing, availability and fault tolerance, to fully exploit flexible resource management schemes on cloud-based computation platforms. Spark Streaming [29], Storm [4], Samza [1], Heron [19] and Flink [5] are the most popular open-source systems providing distributed stream processing and analytics. Big industrial players are also developing in-house distributed stream systems such as Muppet [20], MillWheel [8], Dataflow [9] and StreamScope [21].

**Elasticity.** A large body of work explores the possibility of achieving elasticity. Castro et al. [12] combine the resource re-scaling operation with fault tolerance functionality in distributed stream systems, such that the intermediate states bound with the processing logic are written to persistent storage before migrating to new computation nodes. An adaptive partitioning operator is proposed in Flux [25] to enable partition movement among nodes for load balance. ChronoStream [28] partitions computation states into a collection of fine-grained slice units and dynamically distributes them across nodes to support elasticity. Gedik et al. [17] propose mechanism to scale stateful operators without violating state consistency. However, all existing work achieves elasticity following the resource-centric paradigm which incurs global synchronization and prevents rapid elasticity. Elasticutor avoids the problem by employing a new executor-centric approach. This approach greatly reduces the synchronization overhead in performing workload rebalancing and therefore enables workload redistribution within milliseconds.

**Workload Distribution.** Generic workload distribution for distributed stream systems is a challenging problem, due to the high skew and huge variance in the incoming data stream over time. Shah [25] et al. designed dynamic workload redistribution mechanisms for individual operators in a traditional stream processing framework, e.g. Borealis [7]. Gedik et al. [17] propose a mixed routing strategy to group the workload by its keys to dynamically balance the load in terms of CPU, memory and bandwidth resource. TimeStream [24] adopts a graph restructuring strategy, by directly replacing the original processing topology with a completely new one. Elasticutor not only achieves load balancing in workload distribution, but also takes migration cost minimization and computation locality into consideration.

## 7 Conclusion

We have presented the Elasticutor framework, which enables rapid elasticity for stream processing systems. Elasticutor follows a new executor-centric approach that statically binds executors to operators, but allows executors to scale independently. This approach decouples the scaling of operators from the global synchronization needed for stateful processing. The Elasticutor framework has two building blocks: elastic executors, which perform dynamic load balancing, and a centralized scheduler that optimizes the use of computational resources. Experiments with real-world stock exchange transactions show that, compared with a traditional resource-centric approach to providing elasticity, Elasticutor doubles the throughput and achieves an average latency orders of magnitude lower.

# References

[1] http://samza.apache.org/.

[2] https://github.com/twitter/heron.

[3] http://netty.io.

[4] http://storm.apache.org/.

[5] http://flink.apache.org/.

[6] ABADI, D. J., CARNEY, D., ÇETINTEMEL, U., CHERNIACK, M., CONVEY, C., LEE, S., STONEBRAKER, M., TATBUL, N., AND ZDONIK, S. B. Aurora: a new model and architecture for data stream management. *VLDB J. 12*, 2 (2003), 120–139.

[7] ABADI, D. J., ET AL. The design of the borealis stream processing engine. In *CIDR* (2005), pp. 277–289.

[8] AKIDAU, T., BALIKOV, A., BEKIROĞLU, K., CHERNYAK, S., HABERMAN, J., LAX, R., MCVEETY, S., MILLS, D., NORDSTROM, P., AND WHITTLE, S. Millwheel: fault-tolerant stream processing at internet scale. *Proceedings of the VLDB Endowment 6*, 11 (2013), 1033–1044.

[9] AKIDAU, T., BRADSHAW, R., CHAMBERS, C., CHERNYAK, S., FERNÁNDEZ-MOCTEZUMA, R. J., LAX, R., MCVEETY, S., MILLS, D., PERRY, F., SCHMIDT, E., ET AL. The dataflow model: a practical approach to balancing correctness, latency, and cost in massive-scale, unbounded, out-of-order data processing. *Proceedings of the VLDB Endowment 8*, 12 (2015), 1792–1803.

[10] ARASU, A., BABCOCK, B., BABU, S., DATAR, M., ITO, K., MOTWANI, R., NISHIZAWA, I., SRIVASTAVA, U., THOMAS, D., VARMA, R., AND WIDOM, J. STREAM: the stanford stream data manager. *IEEE Data Eng. Bull. 26*, 1 (2003), 19–26.

[11] CARNEY, D., ÇETINTEMEL, U., RASIN, A., ZDONIK, S., CHERNIACK, M., AND STONEBRAKER, M. Operator scheduling in a data stream manager. In *VLDB* (2003), pp. 838–849.

[12] CASTRO FERNANDEZ, R., MIGLIAVACCA, M., KALYVIANAKI, E., AND PIETZUCH, P. Integrating scale out and fault tolerance in stream processing using operator state management. In *SIGMOD* (2013), pp. 725–736.

[13] CHANDRASEKARAN, S., COOPER, O., DESHPANDE, A., FRANKLIN, M. J., HELLERSTEIN, J. M., HONG, W., KRISHNAMURTHY, S., MADDEN, S. R., REISS, F., AND SHAH, M. A. Telegraphcq: continuous dataflow processing. In *SIGMOD* (2003), pp. 668–668.

[14] DÓSA, G. The tight bound of first fit decreasing bin-packing algorithm is ffd (i) 11/9opt (i)+ 6/9. In *Combinatorics, Algorithms, Probabilistic and Experimental Methodologies*. Springer, 2007, pp. 1–11.

[15] FU, T. Z. J., DING, J., MA, R. T. B., WINSLETT, M., YANG, Y., AND ZHANG, Z. DRS: dynamic resource scheduling for real-time analytics over fast streams. In *ICDCS* (2015), pp. 411–420.

[16] GARY, M. R., AND JOHNSON, D. S. Computers and intractability: A guide to the theory of np-completeness, 1979.

[17] GEDIK, B. Partitioning functions for stateful data parallelism in stream processing. *VLDBJ 23*, 4 (2014), 517–539.

[18] KORF, R. E. Multi-way number partitioning. In *IJCAI* (2009), pp. 538–543.

[19] KULKARNI, S., BHAGAT, N., FU, M., KEDIGEHALLI, V., KELLOGG, C., MITTAL, S., PATEL, J. M., RAMASAMY, K., AND TANEJA, S. Twitter heron: Stream processing at scale. In *SIGMOD* (2015), pp. 239–250.

[20] LAM, W., LIU, L., PRASAD, S., RAJARAMAN, A., VACHERI, Z., AND DOAN, A. Muppet: Mapreduce-style processing of fast data. *Proceedings of the VLDB Endowment 5*, 12 (2012), 1814–1825.

[21] LIN, W., QIAN, Z., XU, J., YANG, S., ZHOU, J., AND ZHOU, L. Streamscope: Continuous reliable distributed processing of big data streams. In *NSDI* (2016), pp. 439–453.

[22] ONGARO, D., RUMBLE, S. M., STUTSMAN, R., OUSTERHOUT, J., AND ROSENBLUM, M. Fast crash recovery in ramcloud. In *Proceedings of the Twenty-Third ACM Symposium on Operating Systems Principles* (2011), ACM, pp. 29–41.

[23] POWERS, D. M. Applications and explanations of zipf's law. In *Proceedings of the joint conferences on new methods in language processing and computational natural language learning* (1998), Association for Computational Linguistics, pp. 151–160.

[24] QIAN, Z., HE, Y., SU, C., WU, Z., ZHU, H., ZHANG, T., ZHOU, L., YU, Y., AND ZHANG, Z. Timestream: reliable stream computation in the cloud. In *EuroSys* (2013), pp. 1–14.

[25] SHAH, M. A., HELLERSTEIN, J. M., CHANDRASEKARAN, S., AND FRANKLIN, M. J. Flux: An adaptive partitioning operator for continuous query systems. In *ICDE* (2003), pp. 25–36.

[26] TAFT, R., MANSOUR, E., SERAFINI, M., DUGGAN, J., ELMORE, A. J., ABOULNAGA, A., PAVLO, A., AND STONEBRAKER, M. E-store: Fine-grained elastic partitioning for distributed transaction processing systems. *Proceedings of the VLDB Endowment 8*, 3 (2014), 245–256.

[27] TIJMS, H. C. *Stochastic modelling and analysis: a computational approach*. John Wiley & Sons, Inc., 1986.

[28] WU, Y., AND TAN, K.-L. Chronostream: Elastic stateful stream computation in the cloud. In *ICDE* (2015), pp. 723–734.

[29] ZAHARIA, M., DAS, T., LI, H., HUNTER, T., SHENKER, S., AND STOICA, I. Discretized streams: Fault-tolerant streaming computation at scale. In *SOSP* (2013), pp. 423–438.